\documentclass[amsmath,amssymb,aps,prd,preprint,groupedaddress]{revtex4-1}

\usepackage{graphicx}
\usepackage{bm}
\usepackage{tensor}

\newcommand*{\dd}{\text{d}}
\newcommand*{\lP}{\ell_p}

\begin{document}

\title{On the detection of primordial gravitational waves produced in bouncing models}

\author{Nelson \surname{Pinto-Neto}}
\email[]{nelsonpn@cbpf.br}
\author{Arthur Scardua}
\email[]{arthur@cbpf.br}
\affiliation{ICRA - Centro Brasileiro de Pesquisas F\'{\i}sicas -- CBPF, \\ rua Xavier Sigaud, 150, Urca,
CEP22290-180, Rio de Janeiro, Brazil}

\date{\today}

\begin{abstract}

It is widely known that bouncing models with a dust hydrodynamical fluid satisfying ${c_s^2=p_d/\rho_d\approx 0}$,
where $c_s, p_d, \rho_d$ are the sound velocity, pressure and energy density of the dust fluid, respectively,
have almost scale invariant spectrum of scalar perturbations and negligible primordial gravitational waves.
We investigate whether adding another fluid with $1/3 < \lambda = p/\rho < 1$, which should dominate near
the bounce, can increase the amplitude of gravitational waves in the high frequency regime, turning them
detectable in near future observations for such range of frequencies. Indeed, we show that the
energy density of primordial gravitational waves is proportional to $k^{2(9\lambda-1)/(1+3\lambda)}$
for wavelengths which become bigger
than the Hubble radius when this extra fluid dominates the background. Hence, as $\lambda \to 1$ (an
almost stiff matter fluid), the energy density of primordial gravitational waves will increase faster
in frequency, turning them potentially detectable at high frequencies. However, there is an extra factor
$I_q(\lambda)$ in the amplitude which decreases exponentially with $\lambda$.
The net effect of these two contributions
turns the energy density of primordial gravitational waves
not sufficiently big at high frequencies in order to be detected by present day or near future observations
for models which satisfy the nucleosynthesis bounds and is symmetric with respect to the bounce. Hence,
symmetric bouncing models where the background is dominated by a dust hydrodynamical fluid with small
sound velocity, do not present any significant
amount of primordial gravitational waves at any frequency range compatible with observations, even
if there are other fields present in the model dominating the bounce phase. Any detection
of such waves will then rule out this kind of models.

\end{abstract}

\pacs{04.30.-w, 98.80.-k, 98.80.Jk}

\maketitle

\section{Introduction}
Cosmological bouncing scenarios solve, by construction, the singularity problem present in
the standard cosmological model~\cite{Novello2008}. As a bonus, they solve other puzzles of the standard
cosmological model, like the horizon and
flatness problems, and they can also supply a mechanism to generate
primordial cosmological perturbations from quantum vacuum fluctuations,
with almost scale invariant spectrum~\cite{Peter2007, Peter2008}, as in
inflationary models~\cite{Turner1997},
when the contracting phase is mainly dominated by a matter fluid (a fluid with
equation of state $p=\lambda\rho$ with $\lambda \approx 0$). Hence, they can also be viewed as alternatives to
inflation, although they are not necessarily contradictory to it.

There are nowadays many mechanisms to generate the bounce, normally they involve
new physics and/or new types of fields. There are
also many open questions and issues to be investigated concerning these
models, for reviews see Refs.~\cite{Novello2008,Battefeld2015}. One of
these questions concerns the presence of primordial gravitational waves. In the case
where the fluid driving the contracting phase is a canonical scalar field,
in which the sound velocity of scalar perturbations $c_s$ is equal to the speed of light, $c_s = c=1$,
the production of primordial gravitational waves is usually very high~\cite{brand},
yielding a tensor to scalar perturbation ratio $r=T/S \approx 1$ which is incompatible
with observations~\cite{Ade2016} ($T$ and $S$ are the amplitudes of tensor and scalar perturbations,
respectively). On the other hand, for K-essence scalar fields, which mimic hydrodynamical fluids and
$c_s = \lambda \approx 0$,
the amplitude of primordial gravitational waves produced is very small~\cite{Bessada2012},
and they cannot be seen in any band of frequency. This feature is compatible
with present cosmological observations, but
it does not offer any testable prediction into which this model could be confronted with future observations.
As it is well known, the detection of gravitational waves emitted by black holes~\cite{PhysRevLett.116.061102}
opened the gravitational waves astronomy era.
One of the possible signals to be detected in different frequency ranges in the next decade,
away from cosmological scales,
are precisely the primordial gravitational waves. Unlike the black hole collision signals
recently detected, these primary waves are stochastic and less intense. The detection of such waves will give
information about the early Universe~\cite{Maggiore2000}, e.g., if there was an inflationary era, a bounce,
or even both.

The aim of this paper is to investigate whether high energy modifications of the
model described in Ref.~\cite{Bessada2012}, a Universe containing radiation and dust which goes
through a quantum bounce, can increase the amplitude of primordial gravitational waves in the high frequency regime. In fact, the energy density of gravitational waves has a spectrum proportional to \[f^{\frac{2(9\lambda_c-1)}{1+3\lambda_c}}~ ,\] where $f$ is the frequency and $\lambda_c$ is the equation of state parameter of the fluid which is dominating the background when the mode is leaving the Hubble radius. Hence, for modes leaving the Hubble radius at the dust dominated phase, it decreases with frequency as $f^{-2}$, and it increases as $f^{2}$ for modes leaving the Hubble radius at the radiation dominated phase. If one adds to the model a stiff fluid with $\lambda \approx 1$, which should dominate its densest phase, so dense that that the sound velocity of the fluid becomes comparable with the speed of light~\cite{Zeldovich1972}, then for modes leaving the Hubble radius at the stiff matter dominated phase, the energy density of gravitational waves would increase with frequency as $f^4$. Our goal is to evaluate whether adding this stiff fluid to the model can sufficiently increase the energy density of gravitational waves in the high frequency regime in a way that they could be detected by future observations, without spoiling the good features of the model (scale invariant spectrum of scalar cosmological perturbations, standard nucleosynthesis phase, etc).

The paper is divided as follows: in the next section we derive the equations for tensor perturbations when the background is quantized, in section III we describe the features of the full background model, and the qualitative features of the evolution of tensor perturbations in this background. In section IV we solve the tensor perturbation equations semi-numerically, and we present analytical approximations in order to understand the numerical results qualitatively. We finish in section V, with the conclusions.

\section{\label{sec:qcosmomodel}Relic gravitons in a quantum bouncing model}

In this section, we first derive the evolution equations for tensor perturbations
in a bouncing cosmological model near the bounce itself, where the background evolution
is dominated by a single perfect fluid with equation of state $p=\lambda \rho$, and
the bounce is caused by quantum effects described in the framework of quantum cosmology
interpreted along the lines of the de Broglie-Bohm quantum theory, see
Refs.~\cite{Bessada2012,Peter2005,Peter2006} for details. Note that the usual Copenhagen point of
view in quantum mechanics cannot be used in quantum cosmology as the whole Universe is being quantized, including
observers themselves, see Ref.~\cite{fabris-nelson} for a review on this subject.
The de Broglie-Bohm quantum theory assumes the existence of an objective reality, where
positions of particles and/or field amplitudes have definite values, independently of any observation.
It is an explicit non-local realistic quantum theory which satisfies all experimental tests already made in quantum
systems. The Bohmian trajectories describing the scale factor evolution in this framework are calculated, and they are usually non-singular, presenting a bounce due to quantum effects at small scales, and turning to a classical standard evolution when the scale factor becomes sufficiently large. Subsequently, we enlarge the model in order to include dust and radiation, which however are not important during
the bounce itself, and are relevant only when the evolution is classical.

The action we consider describing the physics around the bounce contains an Einstein-Hilbert term 
and a single perfect fluid term described by the Schutz formalism~\cite{Schutz,*Schutz2}:
\begin{equation}
 S = S_{\mathrm{GR}} + S_\mathrm{fluid} =
-\frac{1}{6\lP^2} \int \sqrt{-g} R \dd^4 x + \int \sqrt{-g} p\dd^4
x,
\label{action}
\end{equation}
where $\lP=(8\pi\text{G}_N /3)^{1/2}$ is the Planck length ($\hbar=c=1$), 
$p$ is the perfect fluid pressure satisfying $p=\lambda\rho$,
$\rho$ is the fluid energy density, and $\lambda=\mathrm{const}$. The metric
$\bm{g}$ appearing in action~(\ref{action}) describes a background
Friedman-Lema\^{\i}tre-Robertson-Walker (FLRW) metric and a
first-order tensor perturbation $w_{ij}$. It reads
\begin{equation}
\dd s^{2}=N^2\left(\tau\right)\dd\tau^2 -a_{\rm
  phys}^2\left(\tau\right) \left(\gamma_{ij}+w_{ij}\right)\dd x^{i}\dd
x^{j}.
\label{adm2}
\end{equation}

The constant curvature background spacelike metric is given by $\gamma_{ij}$. It lowers and raises
the indices of the tensor perturbation $w_{ij}$, which is
transverse and traceless ($w\indices{^{ij}_{ |j}}=0$ and $w\indices{^i_i}=0$, the bar indicating a covariant derivative with respect
to $\gamma$). $N(\tau)$ is the lapse function, and defines the
time gauge, once fixed. From now on we consider only flat spatial metrics.

Action~(\ref{action}) with metric~(\ref{adm2}) yields, after some suitable
canonical transformations (see Ref.~\cite{Peter2005} for details),
the second-order Hamiltonian
\begin{eqnarray}
H &=& N H_0 \nonumber \\*
 {H}_0 & = &\left[ -\frac{1}{4 {a}} {P}_a^2 +
\frac{ {P}_{_T}}{ {a}^{3\omega}}+\int \dd^3 x \left(
6\frac{ \Pi^{ij} {\Pi}_{ij}}{\gamma^{1/2} a^3}
\frac{1}{24}
\gamma^{1/2} a  {w}_{ij|k} {w}^{ij|k} \right)\right],
\label{h200}
\end{eqnarray}
where $P_a$, ${\Pi}^{ij}$, $ P_{_T}$ are the momenta
canonically conjugate to the scale factor, the tensor
perturbations, and to the fluid degree of freedom, respectively.
These quantities have been redefined in order to be dimensionless, e.g.
$a_{\text phys}=\lP a/\sqrt{V}$, where $V$ is the comoving volume of
the background spacelike hypersurfaces.  The Hamiltonian in Eq.~(\ref{h200})
gives Einstein's equations both at zeroth and first order in perturbation expansion. 
No assumption about the background dynamics has been used in order to arrive
at its final form given in Eq.~(\ref{h200}).

The Dirac quantization of the background and tensor
perturbations can be implemented by imposing
${\hat{H}}_0\Psi(a,w_{ij})=0$, where ${\hat{H}}_0$ is the operator version of
the classical $H_0$ given in Eq.~(\ref{h200}).
The corresponding Wheeler-DeWitt equation is then given by
\begin{widetext}
\begin{eqnarray}
    i\frac{\partial\Psi}{\partial T} & = & \hat{H} a^{3\lambda} \Psi
\equiv\left[ \frac{a^{3\lambda-1}}{4} \frac{\partial^2}{\partial
a^{2}} +  \int d^3 x \left( - 6
\frac{a^{3(\lambda-1)}}{\gamma^{1/2}} \frac{\delta^2}{\delta w_{ij}
\delta w^{ij}} +
a^{3\lambda+1} \gamma^{1/2} \frac{w_{ij|k}
w^{ij|k}}{24}\right)\right]\Psi. 
\label{es2}
\end{eqnarray}
\end{widetext}

We have imposed the time gauge $N=a^{3\lambda}$, yielding $T$ as the time variable. 
Making the separation ansatz for the wave functional
$\Psi[a,w_{ij},T]=\varphi(a,T)\psi[a,w_{ij},T]$, Eq.~(\ref{es2}) can be split into
two, 
\begin{equation}
i\frac{\partial\varphi}{\partial T} =\frac{a^{3\lambda-1}}{4}
\frac{\partial^2\varphi}{\partial a^2} \varphi,
\label{es20}
\end{equation}
and
\begin{eqnarray}
i\frac{\partial\psi}{\partial T}&=& \int \dd^3 x
\left(-6\frac{a^{3(\lambda-1)}}{\gamma^{1/2}}
\frac{\delta^2}{\delta w_{ij} \delta w^{ij}} 
a^{3\lambda+1} \gamma^{1/2} \frac{w_{ij|k}
w^{ij|k}}{24}\right)\psi.
\label{es22}
\end{eqnarray}

Using the de Broglie-Bohm quantum theory~\cite{fabris-nelson}, 
Eq.~(\ref{es20}) can be solved, yielding a Bohmian quantum
trajectory $a(T)$. Using the de Broglie-Bohm framework described in Ref.~\cite{Peter2006}, 
the guidance relation reads
\begin{equation}
\label{guidance} \frac{da}{dT}=-\frac{a^{(3\lambda-1)}}{2}
\frac{\partial S}{\partial a}.
\end{equation}

It is in accordance with the usual Hamilton-Jacobi classical relations $da/dT=\{a,H\}=
-\frac{1}{2}a^{(3\lambda-1)}P_a$ with $P_a=\partial S/\partial a$. Note that
$S$ is the phase of the wave function, and it coincides with the classical action,
yielding the usual Hamilton-Jacobi classical trajectories in the classical limit.

Taking the initial normalized gaussian at $T=0$.
\begin{equation}
\label{initial}
    \Psi^{(\mathrm{init})}(\chi)=\biggl(\frac{8}{T_b\pi}\biggr)^{1/4}
\exp\left(-\frac{\chi^2}{T_b}\right) ,
\end{equation}
where $T_b$ is an arbitrary constant, the solution of Eq.~(\ref{es20}) reads~\cite{Peter2006}
\begin{widetext}
\begin{eqnarray}\label{psi1t}
\Psi(a,T)&=&\left[\frac{8 T_b}{\pi\left(T^2+T_b^2\right)}
\right]^{1/4}
\exp\biggl[\frac{-4T_ba^{3(1-\lambda)}}{9(T^2+T_b^2)(1-\lambda)^2}\biggr]
    \times\nonumber\\*&&\times
\exp\left\{-i\left[\frac{4Ta^{3(1-\lambda)}}{9(T^2+T_b^2)(1-\lambda)^2}
+\frac{1}{2}\arctan\biggl(\frac{T_b}{T}\biggr)-\frac{\pi}{4}\right]\right\},
\end{eqnarray}
\end{widetext}

The phase of this wave solution yields he Bohmian quantum trajectory for the scale factor
\begin{equation}
\label{at} a(T) = a_b
\left[1+\left(\frac{T}{T_b}\right)^2\right]^\frac{1}{3(1-\lambda)},
\end{equation}
where $a_b$ is the scale factor at the
bounce  at $T=0$. Note that this solution has no singularities and
tends to the classical solution when $T\rightarrow\pm\infty$.
The quantity $T_b$, together with $a_b$, gives the curvature scale at the bounce,
$L_{\rm bounce}\equiv T_b a_b^{3\lambda}$.

Once one has $a(T)$ as a prescribed function
of time, one can perform the time dependent unitary transformation
\begin{align}
    U =&\exp\biggl\{ i
\biggl[ \int d^3 x \gamma^{1/2} \frac{a^{\prime} w_{ij} w^{ij}}{2a}
\biggr] \biggr\}\times\nonumber\\*
    &\times\exp\biggl\{ i \biggl[ \int d^3 x \biggl(
\frac{w_{ij}\Pi ^{ij} + \Pi ^{ij} w_{ij}}{2} \biggr)
\ln\biggl( \frac{\sqrt{12}}{a} \biggr) \biggr]\biggr\},
\label{transformacao unitaria}
\end{align}
yielding the following simple form for the Schr\"odinger equation for the perturbations:
\begin{eqnarray}
i\frac{\partial\chi(w,\eta)}{\partial \eta}&=& \int d^3 x
\left\{-\frac{1}{2\gamma^{1/2}} \frac{\delta^2}{{\delta w}^2} +\right.\nonumber\\*
&&\left.+\gamma^{1/2}\left[\frac{1}{2}w_{k} w^{k}
-\frac{a^{\prime\prime}}{2a}w^2\right] \right\} \chi(w,\eta).
\label{simplest}
\end{eqnarray}

We made a transformation to conformal time
$\eta$, $a^{3\omega-1}\dd T = \dd \eta$, and
a prime $'$ denotes the derivative with respect to $\eta$.
This is the same Schr\"odinger equation used in semi-classical
gravity for linear tensor perturbations~\cite{MFB}, but the scale factor which appears
in it is the Bohmian trajectory~(\ref{at}), it is not the classical scale factor.
Remember that Eq.~(\ref{simplest}) was obtained without ever
using the background Einstein's equations. Hence, it can be used when the background
is also quantized, and it can be extended to the classical regime when other fluids
may become relevant. Note that as the Bohmian scale factor $a(\eta)$ given in Eq.~(\ref{at}) approaches
the classical limit after the bounce, the matching of the classical and quantum phases is straightforward.
Note, however, that $a(\eta)$ departs from the classical solution near the bounce, and this
fact leads to some different consequences with
respect to the usual semi-classical approach.

In the Heisenberg representation, the equations for the operator
${\hat{w}}_{ij}$ read
\begin{equation}
{\hat{w}}_{ij}''+2\frac{a'}{a}{\hat{w}}_{ij}' -
{\hat{w}}_{ij|k}^{|k}
 = 0,
\label{tenspereq}
\end{equation}
which corresponds to the usual equation for quantum tensor perturbations
in classical backgrounds~\cite{MFB}.

It is convenient to expand these quantum mechanical operators into Fourier
modes and subject them to quantization rules:
\begin{eqnarray}
 {\hat{w}}_{ij} \left(x\right) & = & \sqrt{6}\lP
\sum_{\alpha = +,\times}\int \frac{d^3
k}{\left(2\pi\right)^{3/2}}
\varepsilon^{(\alpha)}_{ij}\left[w^{(\alpha)}_k\left(\eta\right)
e^{-i\mathbf{k}\cdot\mathbf{x}}\hat{a}_\mathbf{k}^{(\alpha)}\right.\nonumber
\\* & + & \left.w^{\ast(\alpha)}_k\left(\eta\right)
e^{i\mathbf{k}\cdot\mathbf{x}}\hat{a}_\mathbf{k}^{(\alpha)\dagger}
\right],
\label{expw}
\end{eqnarray}
where $x=\left(\eta,\mathbf{x}\right)$,
$\varepsilon^{(\alpha)}_{ij}=\varepsilon^{(\alpha)}_{ij}\left(\hat{\mathbf{k}}\right)$
is the polarization tensor for the two graviton polarization
states $+$ and $\times$ labeled by $\alpha$, and satisfies
\begin{equation}
\varepsilon^{(\alpha)ij}\varepsilon^{(\alpha')}_{ij}=2\delta_{\alpha\alpha'}.
\label{polten}
\end{equation}

Also, $w^{(\alpha)}_k\left(\eta\right)$ are mode functions, and
$\hat{a}_\mathbf{k}^{(\alpha)\dagger}$,
$\hat{a}_\mathbf{k}^{(\alpha)}$ are the usual creation and annihilation
operators, respectively. Such operators satisfy the equal-time
commutation relations
\begin{eqnarray}
\label{comma1}
&&\left[\hat{a}_\mathbf{k}^{(\alpha)},\hat{a}_\mathbf{k'}^{(\alpha')\dagger}\right]
= \delta_{\alpha\alpha'}
\delta^{(3)}\left(\mathbf{k}-\mathbf{k'}\right),\\*
&&\left[\hat{a}_\mathbf{k}^{(\alpha)},\hat{a}_\mathbf{k'}^{(\alpha')}\right]
=
\left[\hat{a}_\mathbf{k}^{(\alpha)\dagger},\hat{a}_\mathbf{k'}^{(\alpha')\dagger}\right]
= 0,
\label{comma2}
\end{eqnarray}
and the quantum vacuum is defined by
\begin{equation}
\label{defvac} \hat{a}_\mathbf{k}^{(\alpha)}| 0 \rangle = 0.
\end{equation}

Inserting the above Fourier expansion into Eq.~(\ref{tenspereq}),
we obtain the mode equation
\begin{equation}
\label{modeeq} w^{(\alpha)\prime\prime}_k +
2\frac{a'}{a}w^{(\alpha)\prime}_k + k^2
w^{(\alpha)}_k = 0.
\end{equation}

Introducing the canonical amplitude $v^{(\alpha)}_k$ as
\begin{equation}
 v^{(\alpha)}_k\equiv a w^{(\alpha)}_k,
\end{equation}
the mode equation~(\ref{modeeq}) becomes
\begin{equation}
 v_k^{(\alpha)''}+ \left(k^2
-\frac{a''}{a} \right)v^{(\alpha)}_k =0,
\label{modeequation}
\end{equation}
for each graviton polarization state. From now on, we will omit the index $\alpha$.
We will also impose vacuum initial conditions when $\eta\to - \infty$ and $a\to \infty$)
\begin{equation}
v_k(\eta\to-\infty)=\frac{e^{-i k \eta}}{\sqrt{2 k}}
\label{eq:initialcond}
\end{equation}
One is now able to evolve the gravitational wave mode equation from the initial condition~(\ref{eq:initialcond}) to its amplitude today. The quantum and classical behaviors will impact over the evolution through the potential $a''/a$.

In the background model studied in Ref.~\cite{Bessada2012}, the fluid dominating at the
bounce was radiation, $\lambda=1/3$, with an additional
dust fluid dominating when the Universe was large, in order to furnish an almost scale invariant spectrum of
scalar cosmological perturbations. As explained in the introduction, we will now investigate the situation where
the bounce is dominated by an extra fluid with $1/3 < \lambda < 1$, together with dust and radiation,
which are not relevant
near the bounce. In the next section we will describe the full background model and its qualitative features.

\section{The full background model}

The present model contains three non-interacting perfect fluids: dust, radiation, and a fluid satisfying $p=\lambda \rho$, with $1/3 < \lambda < 1$, usually with $\lambda \approx 1$, which we call almost stiff matter (asm). The dust fluid controls the dynamics of the Universe when it is large, and the asm dominates its dynamics near the bounce, when the curvature scalar reaches its highest values\footnote{The curvature scale is proportional to the inverse of the square root of the curvature scalar.}, and the Universe moves from the contracting to the expanding phase. The radiation fluid dominates in between these two fluids. When the curvature scale approaches the Planck length scale, the scale factor gets near its smallest value $a_b$, and quantum effects realize the transition between contraction to expansion, the bounce. This quantum phase is dominated by the asm fluid.

The radiation and dust fluid model massless or ultra-relativistic massive fields, and cold massive fields, respectively. The asm fluid can represent the content of the Universe when it was so dense
that the sound velocity of the fluid becomes comparable with the speed of light \cite{Zeldovich1972}.

In order to satisfy cosmological observations and the model hypotheses, there are some constraints the asm fluid must fulfill\footnote{Imposing these constraints will limit the amplification of gravitational waves in the asm era, as we will see.}:
\begin{itemize}
	\item The quantum effects must be restricted to the asm dominated phase;
	\item Radiation must dominate during nucleosynthesis;
	\item There must be a classical region between asm and radiation.
\end{itemize}

\begin{figure}
\centering
\includegraphics[width=\textwidth]{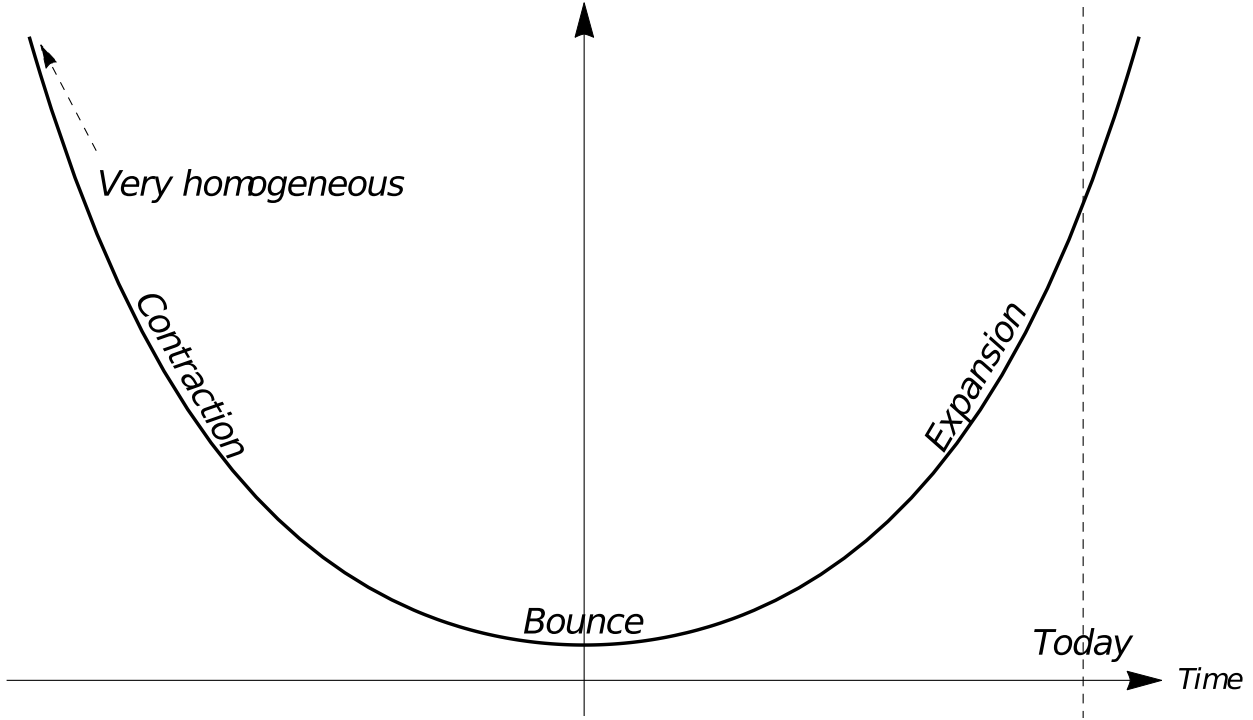}
\caption{\label{fig:scale_factor}Evolution of the scale factor in parametric time.}
\end{figure}

As shown in Fig.~(\ref{fig:scale_factor}), the Universe had a contracting phase in the past, when it was almost flat and very homogeneous. The inhomogeneities were generated by quantum vacuum fluctuations at this phase, and amplified afterwards. The tensorial quantum stochastic fluctuations generated in this contracting past were the sources of the primordial stochastic gravitational waves which could be observed today\footnote{As they are stochastic, there is no coherent time-dependent signal
that could be detected using a match-filtering method as used in the first direct detection of gravitational waves~\cite{PhysRevLett.116.061102}}.

\begin{figure}
\centering
\includegraphics[width=\textwidth]{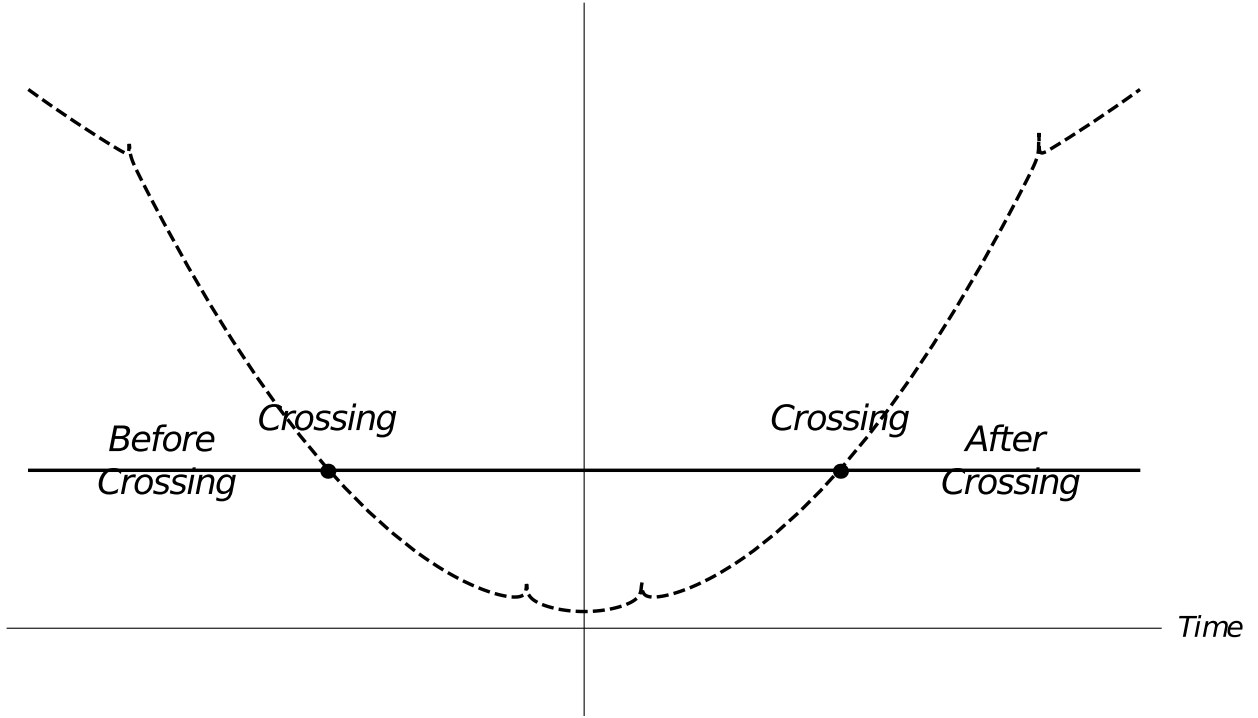}
\caption{\label{fig:I}Crossing the curvature scale. The co-moving wavelength $\lambda=1/k$, the horizontal line, is smaller than the co-moving curvature scale $|a/(a'')|^{1/2}$, the dashed line, in the far past and in the far future of the history of the Universe, when the tensor mode oscillates, and it becomes bigger around the bounce, when the tensor mode gets amplified.}
\end{figure}

Waves with different frequencies will have different amplifications, depending when their wavelengths becomes bigger than the Universe curvature scale. When they are smaller, they do not feel the curvature of the Universe and they oscillate as free fields in flat space-time. When their wavelengths become bigger than the curvature scale, they are pumped by the gravitational field, and they get amplified. Fig.~(\ref{fig:I}) shows a comparison between the co-moving wavelength $\lambda=1/k$ and the co-moving curvature scale $|a/(a'')|^{1/2}$ along the history of the Universe.

This amplification changes
according to which fluid dominates the dynamics of the background when the crossing occurs. Hence, we
expect to obtain different dependence of amplitude with frequency for each different fluid domination.

The background model have two regimes. A classical and a quantum regime. In the classical one, the Friedmann equation relates the scale factor $a$ and the conformal time $\eta$ through the equation
\begin{equation}
\dot{a}= \text{Sign}(\eta) H_0 \sqrt{\Omega_r+\Omega_d a+\Omega_\lambda a^{(1-3\lambda)}} ,
\label{eq:friedmann1}
\end{equation}
where $\Omega_i \equiv \rho_i/\rho_c$, $i=r,d,\lambda$ and $\rho_c$ is the critical density today; $H_0$ is the Hubble factor today; and $\lambda$ is the fluid parameter of asm, i.e., $p_{asm}=\lambda \rho_{asm}$. We set
$a_{\rm today} \equiv a_0=1$.

The critical densities must satisfy the constraints of observation: the equality between radiation and dust must occur in the redshift $2740$, and asm must dominate earlier than the nucleosynthesis era, which occurs at redshift $10^7$~\cite{Mukhanov200512}:
\begin{align}
\Omega_r&=\Omega_d \frac{1}{1+z_e}\\*
 \Omega_r &>\Omega_\lambda \left(\frac{1}{1+z_n}\right).
 \label{eq:rest1}
\end{align}

In the quantum regime, Eq.~(\ref{at}) yields
\begin{equation}
	\dot{a}=\text{Sign}(\eta) H_0 \sqrt{\Omega_\lambda a^{1-3\lambda}\left[1-\left(\frac{a_b}{a}\right)^{3(1-\lambda)}\right]}.
    \label{eq:friedq}
\end{equation}

There is a period when both Eq.~(\ref{eq:friedmann1}) and Eq.~(\ref{eq:friedq}) are valid, dominated by a classical asm, which happens when \[\left(\frac{a_b}{a}\right)^{3(1-\lambda)} \ll 1.\]
Let us take $\left(\frac{a_b}{a}\right)^{3(1-\lambda)}<\frac{1}{100}\ll 1$. Equality between asm and radiation happens for the scale factor \[ \left(\frac{\Omega_{\lambda}}{\Omega_r}\right)^\frac{1}{3\lambda-1}.\] Then we get,
\begin{equation}
a_b 10^\frac{2}{3(1-\lambda)} < a < \left(\frac{\Omega_{\lambda}}{\Omega_r}\right)^\frac{1}{3\lambda-1} < a_n,
\label{eq:constrain}
\end{equation}
where $a_n$ is the scale factor at the nucleosynthesis era.
Equation~(\ref{eq:constrain}) constrains $\Omega_{\lambda}$ with respect to the scale factor in the bounce $a_b$, and the fluid parameter $\lambda$. Because of this equation, the stiffness of the fluid is limited to
\begin{equation}
\lambda < 1- \frac{2}{3 \text{Log}_{10} \left(\frac{a_n}{a_b}\right)}.
\label{eq:omegaconstrain}
\end{equation}

The amplitude of gravitational waves satisfies the wave Eq.~(\ref{modeequation})
\begin{equation}
 v_k^{''}+\left( k^2
-\frac{a''}{a} \right)v_k =0,
\label{modeequation2}
\end{equation}
where the potential takes the form
\begin{align}
\frac{a''}{a}=\frac{H_0^2}{2}\left[\frac{\Omega_d}{a}-\left(3\lambda-1\right)\frac{\Omega_\lambda}{a^{3\lambda+1}}\right]&\text{Classical}\\*
\frac{a''}{a}=\alpha^2 \left(\frac{ab}{a}\right)^4\left[1-\frac{3\lambda-1}{2}\left(\frac{a_b}{a}\right)^{3(\lambda-1)}\right]&\text{Quantum},
\end{align}
where\[\alpha^2\equiv \frac{H_0^2 \Omega_{\lambda}}{a_b^{1+3\lambda}}\]

\begin{figure}
	\includegraphics[width=\textwidth]{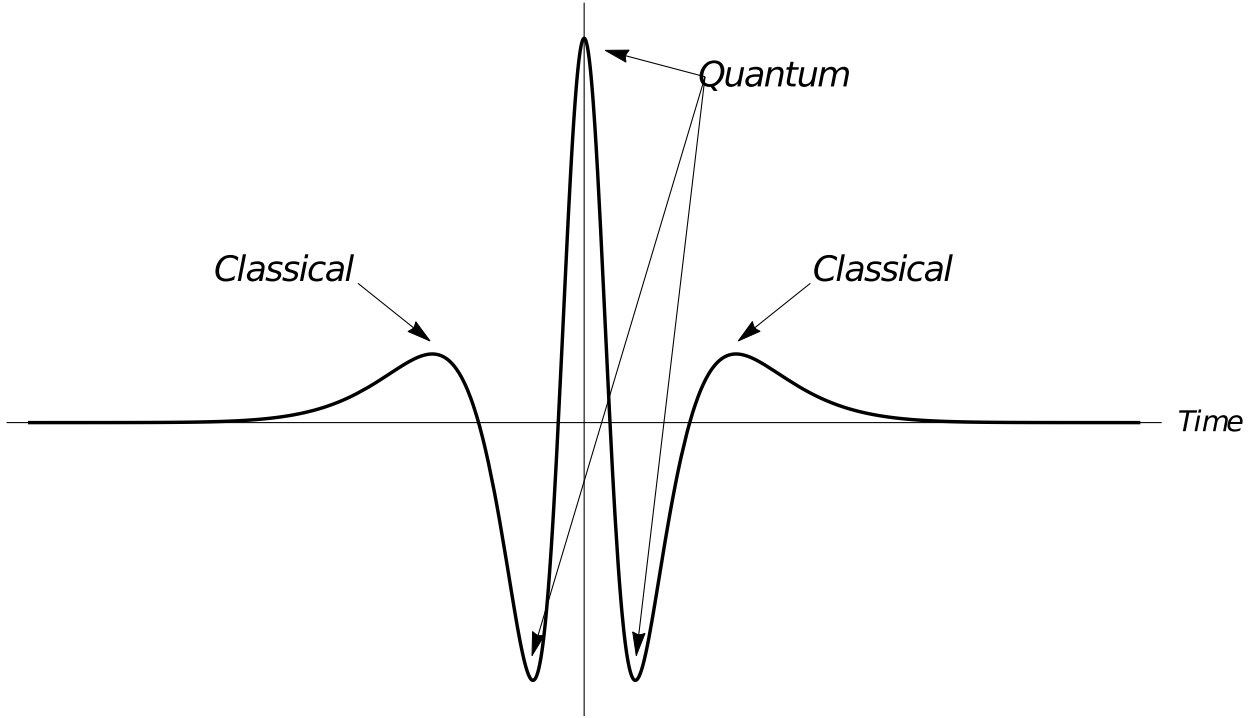}
	\caption{\label{fig:Potential}Structure of the potential $a''/a$ (not in scale).}
\end{figure}

The behavior of the potential is shown in Fig.~(\ref{fig:Potential}). Two maxima are classical due to the transition radiation-asm, one in each bounce side. The two minima come from the quantum regime, and the highest peak happens in the bounce.

\section{Numerical solutions and analytical approximations}

\begin{figure}
	\includegraphics[width=\textwidth]{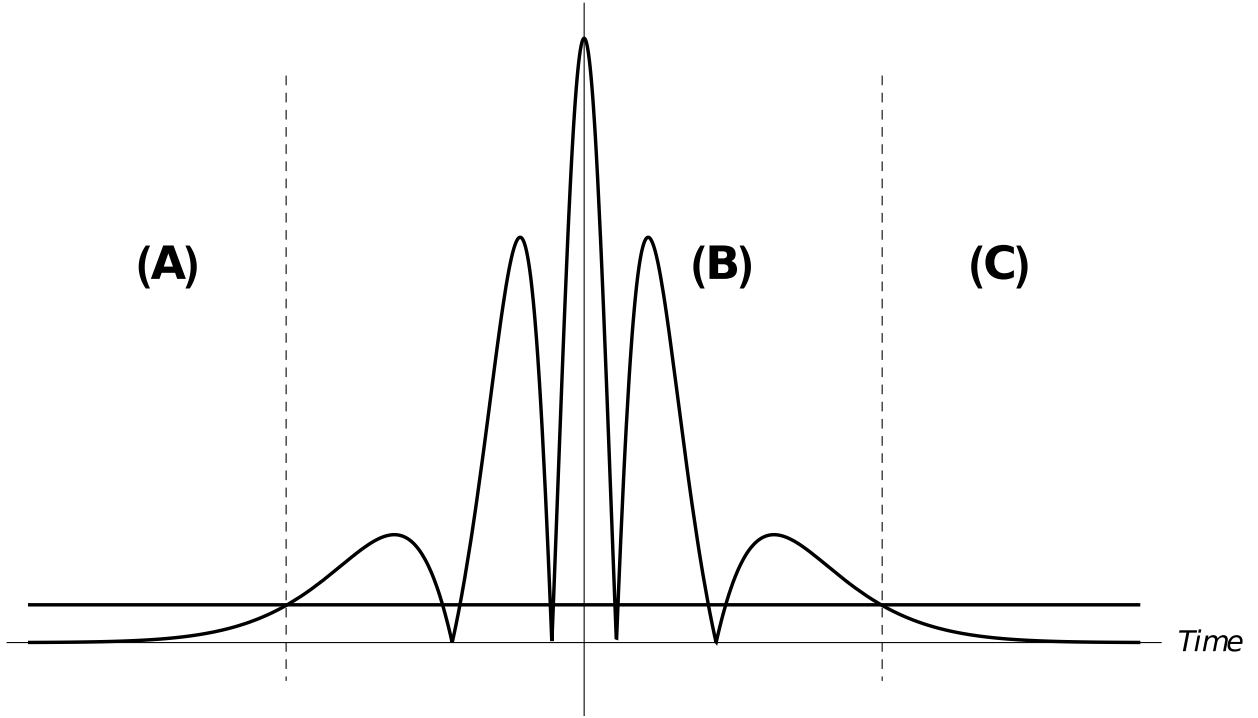}
	\caption{\label{fig:crossing}Crossing the potential $a''/a$ (it is not in scale).}
\end{figure}

For a better understanding on how the different fluids present in the model control the amplitude of gravitational waves, it is necessary a semi-analytical approach. Such approximation can be done separating the evolution in three regions, as shown in Fig.~\ref{fig:crossing}.

\begin{itemize}
	\item[(A)] Outside the potential, or inside the curvature scale: $k\gg \frac{a''}{a}$
	\item[(B)] Inside the potential, or outside the curvature scale: $k\ll \frac{a''}{a}$
	\item[(C)] Outside the potential again, or re-entering the curvature scale: $k\gg \frac{a''}{a}$
\end{itemize}

There are regions in B where $k>\frac{a''}{a}$, but they are negligible.

In A and C, the solutions are oscillatory. Using the quantum initial condition given in Eq.~\eqref{eq:initialcond}, we get
\begin{align}
v(\eta)&=\frac{e^{-i k \eta}}{\sqrt{2k}}, \qquad & \text{in (A)}\label{eq:ini1}\\*
v(\eta)&=C_1 e^{-ik\eta}+C_2 e^{i k \eta}, \qquad & \text{in (C)}.
\end{align}

In (B), the zero order term neglecting $k$ reads
\begin{equation}
v(\eta)=a(\eta) \left[B_1+B_2 \int^{\eta}_{-\eta_c}\frac{d\bar\eta}{a^2 (\bar\eta)} \right],
\label{eq:ampli1}
\end{equation}
where $-\eta_c$ denotes the conformal time when $k^2=\left|\frac{a''}{a}\right|$ in the contracting phase, $\eta=0$ is the bounce conformal time, and $\eta_c$ is the conformal time when the solution exits the potential again (the potential  $a''/a$ is symmetric). The constants can be obtained through matching conditions, and read,
\begin{align}
B_2&=(av'-va')|_{-\eta_c}\label{eq:B2}\\*
B_1&=\frac{v(-\eta_c)}{a(-\eta_c)}.
\label{eq:B1}
\end{align}

From now on, ${a_c=a(\eta_c)=a(-\eta_c)}$ and ${a'_c=|a'(\eta_c)|=a'(\eta_c)=-a'(-\eta_c)}$.

The constants in Eq.~(\ref{eq:B2}) and Eq.~(\ref{eq:B1}) are, using Eq.~(\ref{eq:ini1})
\begin{align}
B_1&=\frac{e^{-ik\eta_c}}{a_c \sqrt{2k}}\\*
B_2&=\frac{e^{ik\eta_c}}{a_c\sqrt{2k}}\left(a'_c a_c-ik a_c^2\right).
\end{align}

Therefore, Eq.~(\ref{eq:ampli1}) can be expressed as
\begin{equation}
v(\eta_c)=\frac{e^{i k \eta_c}}{\sqrt{2k}}\left[1+\left(a'_c a_c-ik a_c^2\right) I(a_c)\right],
\end{equation}
where
\begin{equation}
I(a_c)=\int_{-\eta_c}^{\eta_c} \frac{d\eta}{a^2(\eta)}=2 \int^{a_c}_{a_b}\frac{da}{a^2 |a'(a)|}.
\label{Ic}
\end{equation}

Using the fact that $B_2$ is constant, the derivative in the region B can be expressed as
\begin{align}
v'&=\frac{B_2}{a}+v \frac{a'}{a}\notag\\*
\Rightarrow v'(\eta_c)&=\frac{e^{ik\eta_c}}{\sqrt{2k}}\left[\frac{a'_c}{a_c}\left(2+a'_c a_c I(a_c)\right)-ik\left(1+a'_c a_c I(a_c)\right)\right].
\label{eq:integral}
\end{align}

With the functions $v$ and $v'$ in region B determined, the constants present in the function in region C are
\begin{align}
C_1&=\left[v(\eta_c)+\frac{v'(\eta_c)}{-ik}\right] \frac{e^{i k \eta_c}}{2}\\*
C_2&=\left[v(\eta_c)-\frac{v'(\eta_c)}{-ik}\right] \frac{e^{-i k \eta_c}}{2}.
\end{align}

The critical energy of gravitational waves~\cite{Maggiore2000} when the waves reenter the curvature scale
is then given by,
\begin{align}
\Omega_g &\simeq \frac{k^5 l_p^2}{3 \pi^2 H_0^2}\left(|v|^2+\left|\frac{v'}{k}\right|^2\right)=
\frac{2k^5 l_p^2}{3 \pi^2 H_0^2}(|C_1|^2 + |C_2|^2) \notag\\*
	 &=\frac{k^4 l_p^2}{3 \pi^2 H_0^2}\left[ 2+ 4 a'_c a_c I(a_c)+ a'^2_c a^2_c I^2(a_c)+k^2 a_c^4 I^2(a_c)+\right.\nonumber\\*
     &\left.+\frac{a_c^{'2}}{a_c^2 k^2}\left(4+4a'_c a_c I(a_c)+a^{'2}_c a_c^2I^2(a_c)\right)\right].
     \label{eq:omega}
\end{align}

The peak of the potential, which happens at the bounce, leads to a maximum $k$
\begin{equation}
k^2_M = \frac{3(1-\lambda)}{2} \alpha^2 \Rightarrow \frac{k^2_M}{H_0^2} =
\frac{3(1-\lambda)\Omega_{\lambda}}{2 a_b^{1+3\lambda}}.
\end{equation}

As $10^{-31} < a_b \ll 10^{-11}$ (see Ref.~\cite{diogo} for an estimation on that,
remembering that we are setting $a_0=1$),
this is a huge physical frequency, and implies a minimum
physical wavelength many orders of magnitude smaller than the Hubble radius today.
For frequencies smaller than this huge maximum frequency,
the term $I^2(a_c)$ dominates in Eq.~(\ref{eq:omega}). In fact, as the integrand in Eq.~\eqref{Ic}
is a decreasing function of $a$, one has
\begin{equation}
a_c |a'_c| I(a_c) = 2a_c |a'_c|\int_{a_b}^{a_c} \frac{ da}{a^2 |a'|}
\gg 2 a_c |a'_c|\frac{(a_c- a_b)}{a_c^2 |a'_c|}\simeq 2,
\label{eq:intgreater}
\end{equation}
when $a_c \gg a_b$, which is the case for $k\ll k_M$. As in the crossing
$a''_c/a_c \simeq (a'c/a_c)^2 \simeq k^2$, and as
\begin{equation}
\label{Iq}
I(a_c) = 2\int_{a_b}^{a_c} \frac{ da}{a^2 |a'|} \simeq 2\int_{a_b}^{a_q} \frac{ da}{a^2 |a'|} \equiv I_q,
\end{equation}
because the integrand in $I(a_c)$ is dominated by small values of $a$ ($a_q$ denotes the scale factor in the beginning of the quantum phase), the energy density can be expressed as
\begin{equation}
\Omega_g\propto \frac{k^6 l_p^2}{3 \pi^2 H_0^2}I^2_q a_c^4,
\label{eq:appOmega}
\end{equation}
nothing that $I_q$ does not depend on $a_c$. As
\[a_c\approx\left(\frac{H_0^2 \Omega_\lambda}{k^2}\right)^\frac{1}{1+3\lambda_c},\] we obtain
\begin{align}
\Omega_g &\propto \frac{ l^2_p}{3 \pi^2} I^2_q \left(\frac{\Omega_{\lambda_c}}{2}\right)^\frac{4}{1+3\lambda_c} \biggl(\frac{k}{H_0}\biggr)^\frac{2(9\lambda_c-1)}{1+3\lambda_c}\notag\\*
&\propto  k^\frac{2(9\lambda_c-1)}{1+3\lambda_c},
\label{eq:omegaBehave}
\end{align}
where $\lambda_c$ is the equation of state parameter of the fluid which is dominating the background when the mode is leaving the Hubble radius\footnote{In a cosmological model described by general relativity
with single fluid domination, leaving the
Hubble radius is the same as leaving the curvature scale and as crossing the potential}.

Equation~(\ref{eq:omegaBehave}) shows that frequencies that crosses the potential in the dust era ($\lambda=0$) have energy density decaying with $f^{-2}$; the ones entering the potential in the radiation era have energy density growing with $f^{2}$; and frequencies that crosses the potential in the asm era have energy density growing with $f^{4}$.
For frequencies $k \geq k_M$, the integral $I_c$ is zero, since the waves never crosses the curvature scale. In this case, Eq.~\eqref{eq:omega} is dominated by the first term inside the braces, and hence the energy density grows also
as $f^{4}$. It is the usual flat spacetime ultraviolet divergence. These behaviors are shown
in Figs.~(\ref{fig:amplitude4low},\ref{fig:stateparameter},\ref{fig:energydensitychange}) below.

Concerning the amplitudes, the term which contributes mostly to the energy density is the quantum part of the integral Eq.~(\ref{eq:integral}):
\begin{align}\label{eq:intquant}
	I_q &= \int_{a_b}^{a_q} \frac{da}{a^2 |a'|} = \frac{1}{\alpha} \int^{a_q}_{a_b}\frac{da}{a^2 \sqrt{\left(\frac{a_b}{a}\right)^{3\lambda-1}-\left(\frac{a_b}{a}\right)^2}} \notag\\*
		   &= \frac{2}{H_0 \sqrt{\Omega_\lambda a_b^{3(1-\lambda)}}} \frac{\arctan\left(\sqrt{\left(\frac{a_q}{a_b}\right)^{3(1-\lambda)}-1}\right)}{3(1-\lambda)}.
\end{align}

\begin{figure}
	\includegraphics[width=\textwidth]{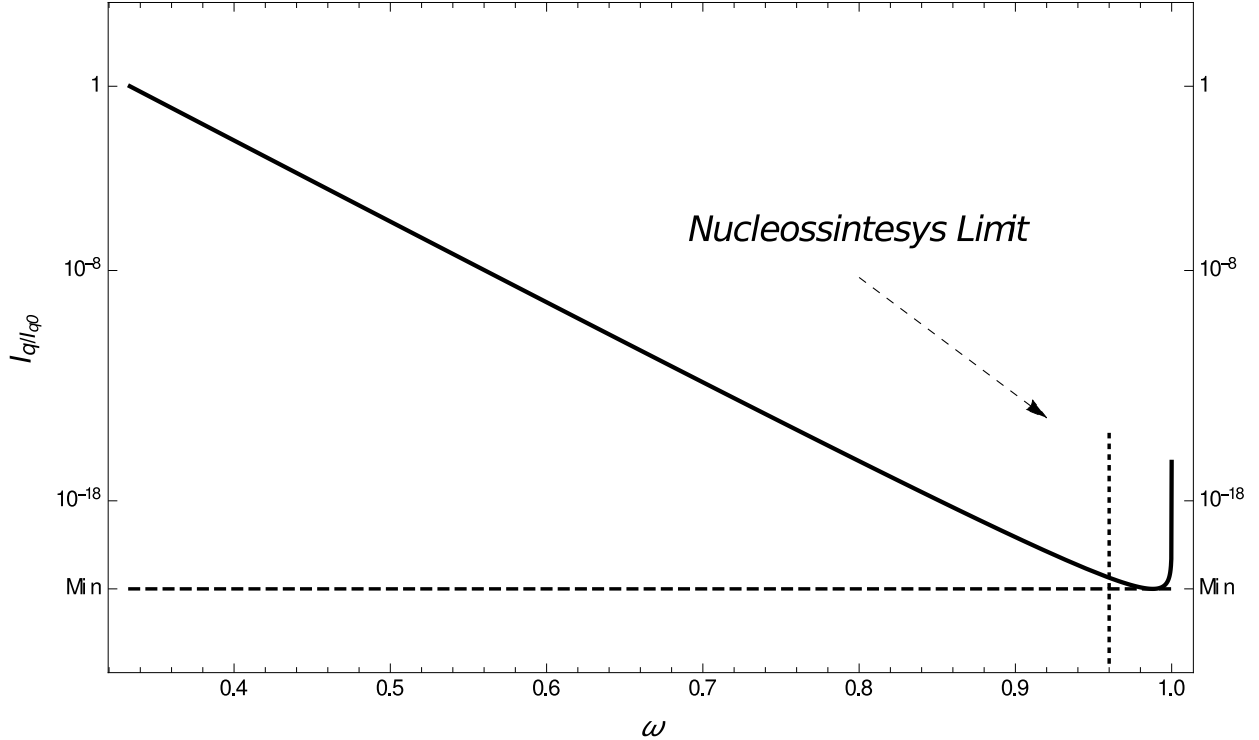}
	\caption{\label{fig:omegaint}The ratio between the integral Eq.~(\ref{eq:intquant}) and its value for $\lambda=1/3$ for different $\lambda$, considering $a_b=10^{-24}$ and $(a_q/a_b)^{3(1-\lambda)}=100$. The minimum value is when $\lambda\approx 0.99$. However, due to the constraint Eq.~(\ref{eq:omegaconstrain}), $\lambda$ is limited to $\lambda\approx 0.96$.}
\end{figure}

Its dependency on $\lambda$ shows that it decreases until ${\lambda\approx 1+\frac{2}{3 \ln (a_b)}}$, when it reaches its minimum value, then it increases rapidly to infinity, when $\lambda=1$, as shown in Fig~\ref{fig:omegaint}.

However, $\lambda$ is limited to the constraint Eq.~(\ref{eq:omegaconstrain}), which is also indicated
in Fig~\ref{fig:omegaint}. It shows that, although the energy density increases more in frequency for higher values
of $\lambda$ as shown in Eq.~\eqref{eq:omegaBehave}, the value of $I_q$ decreases significantly with $\lambda$ in
its physical allowed region, as shown in Fig~\ref{fig:omegaint}. The combination of these two behaviors implies a net decreasing in the amplitude with respect to the case without the asm fluid, as shown in
Fig.~\ref{fig:amplitude4low}.
\begin{figure}
	\includegraphics[width=\textwidth]{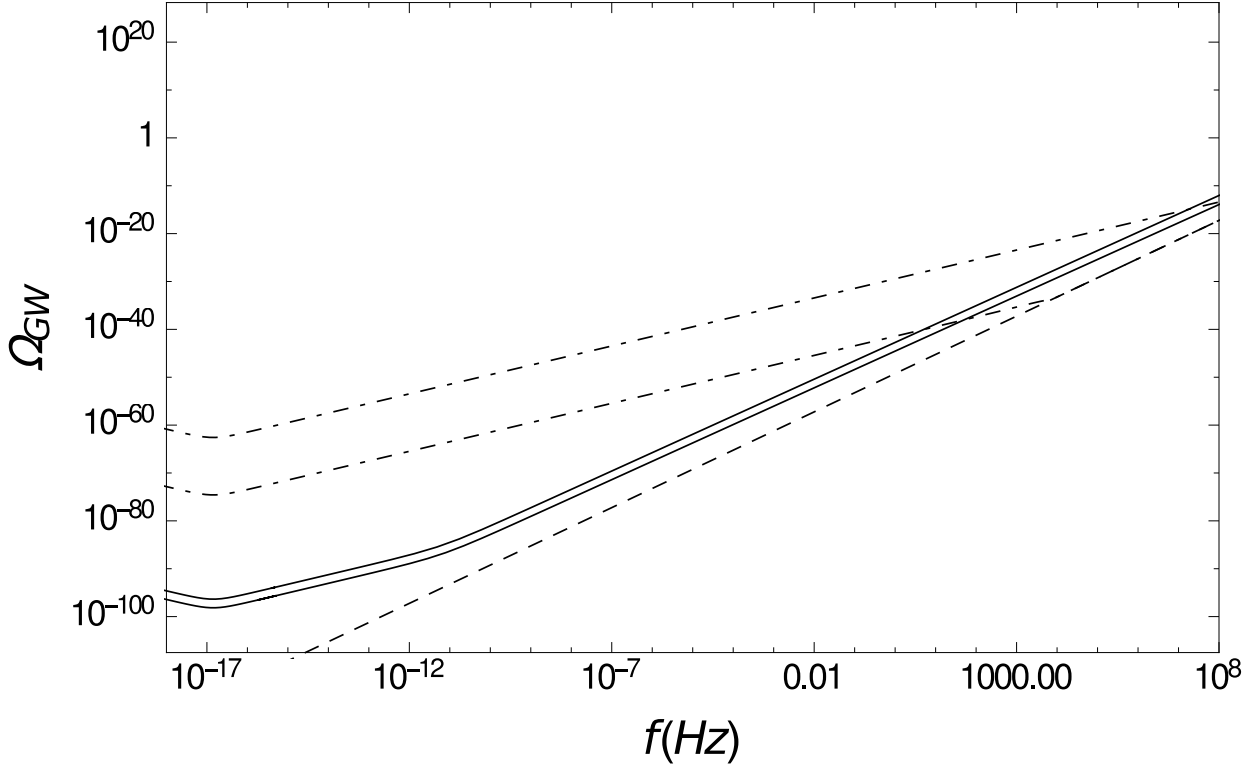}
	\caption{\label{fig:amplitude4low}The energy density of gravitational waves with state parameter $\lambda=1/3$ and minimum scale factors $a_b=10^{-24},10^{-18}$, represented by dot-dashed curves, and state parameter $\lambda=0.9$ with minimum scale factors $a_b=10^{-24},10^{-30}$, represented by continuous curves. The higher energy densities correspond to smaller $a_b$, respectively. The dashed curve corresponds to the limit where the frequency never enters the potential. The different inclinations of the curves are in accordance with the discussion after Eq.~\eqref{eq:omegaBehave}.}
\end{figure}

Note that the usual increasing in the energy density due to the depth of the bounce is quite suppressed due to the presence of the asm fluid. Indeed, the ratio between different gravitational waves energy densities for two different bouncing models with
different scale factors at the bounce, $a_{b1}$ and $a_{b2}$, reads, using Eq.~(\ref{eq:intquant}),
\begin{equation} 
	\frac{\Omega_{g1}}{\Omega_{g2}} = \left(\frac{a_{b2}}{a_{b1}}\right)^{3(1-\lambda)} \frac{\arctan\left(\sqrt{\left(\frac{a_q}{a_{b1}}\right)^{3(1-\lambda)}-1}\right)}{\arctan\left(\sqrt{\left(\frac{a_q}{a_{b_2}}\right)^{3(1-\lambda)}-1}\right)}.
\label{eq:bouncedeep}
\end{equation}

Hence, for fluids with state parameter close to $1$ dominating during the bounce, the increase in intensity due to the bounce depth is exponentially suppressed, as shown in Fig.~\ref{fig:omegaint}.

Let us now summarize the properties inferred by our analytical approximations, and exhibit them with numerical calculations:

\begin{itemize}
	\item The energy density of primordial gravitational waves decreases with the energy density of the fluid which dominates at the bounce. See equations Eq.~(\ref{eq:omegaBehave}) and Eq.~(\ref{eq:intquant}), together with
Fig.~\ref{fig:energydensitychange}.
\begin{figure}
	\includegraphics[width=\textwidth]{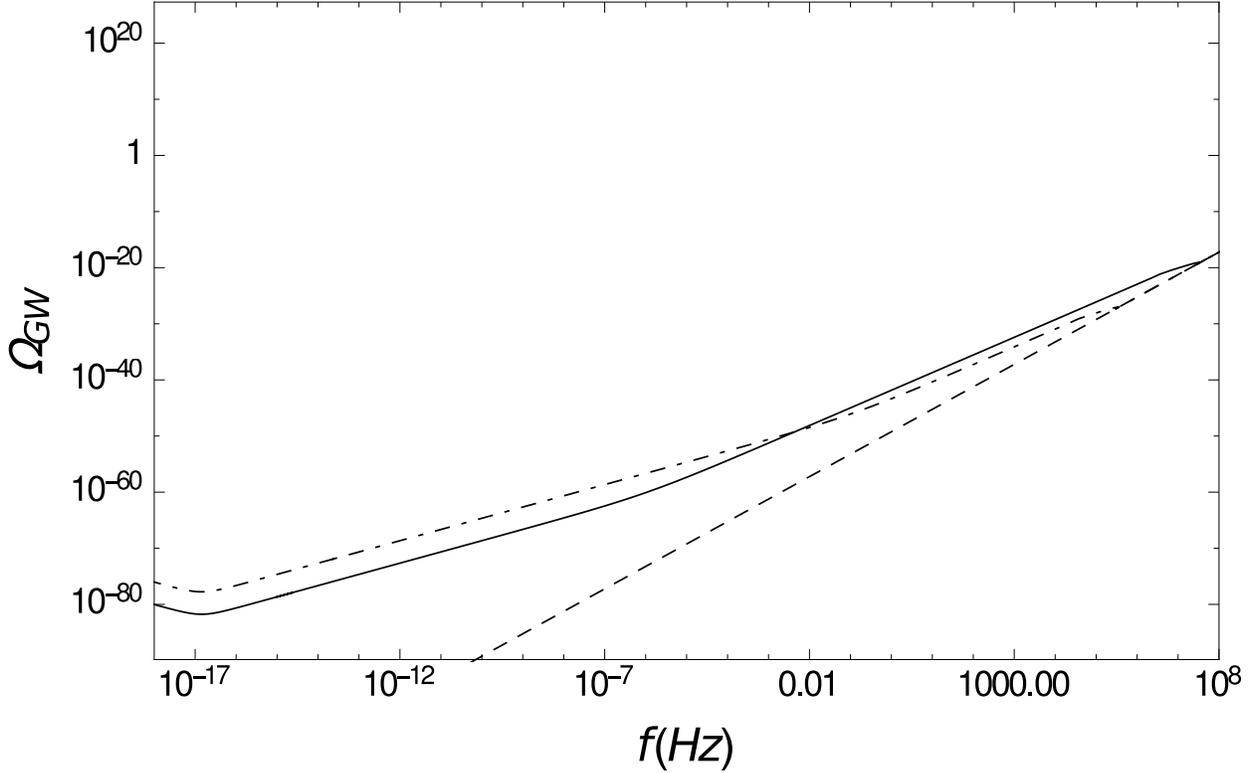}
	\caption{\label{fig:energydensitychange}Gravitational waves energy density dependence on $\Omega_\lambda$ for minimum scale factor $a_b=10^{-24}$, and ams fluid parameter $\lambda=0.6$. The dot-dashed curve corresponds to $\Omega_\lambda=10^{-19}$, while the continuous curve corresponds to $\Omega_\lambda=10^{-15}$. The dashed curve corresponds to the limit where the frequency never enters in the potential. Again, the different inclinations of the curves are in accordance with the discussion after Eq.~\eqref{eq:omegaBehave}.}
\end{figure}

\item The increasing of the energy density of primordial gravitational waves in frequency for increasing
$\lambda$ with $1/3 < \lambda <1$ does not usually compensate the decreasing of its intensity due to the decreasing of $I_q$ with $\lambda$ presented in Fig.~\ref{fig:omegaint}. This compensation usually happens only for very high frequencies, inaccessible by nowadays experiments, see Fig.~\ref{fig:stateparameter}.
\begin{figure}
	\includegraphics[width=\textwidth]{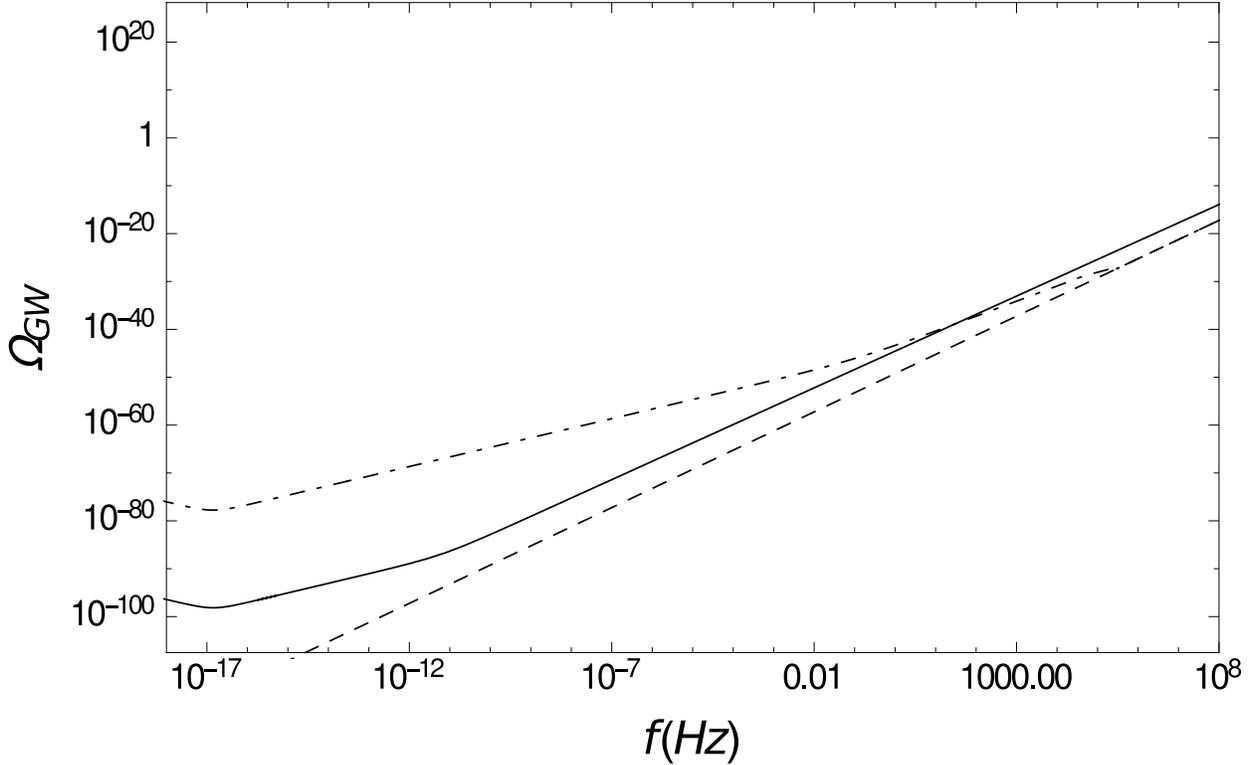}
	\caption{\label{fig:stateparameter}Behavior of the energy density of primordial gravitational waves with respect to the equation of state parameter $\lambda$ for minimum scale factor $a_b=10^{-24}$, and ams energy density today given by $\Omega_\lambda=10^{-19}$. The fluid parameter $\lambda=0.9$ corresponds to the continuous line, and $\lambda=0.6$, corresponds to the dot-dashed line. The dashed curve correspond to the limit where the frequency never enters in the potential. Once again, the different inclinations of the curves are in accordance with the discussion after Eq.~\eqref{eq:omegaBehave}.}
\end{figure}

\item The energy density of primordial gravitational waves is more sensitive to the depth of the bounce for lower equation of state parameters $\lambda$, as shown in the Eq.~\eqref{eq:bouncedeep}. This sensitivity is shown in Fig.~\ref{fig:amplitude4low}.

\item Finally, Fig.~\ref{fig:High} presents one of the highest energy densities of primordial gravitational waves we found for one particular bouncing model, comparing it with results from inflation and present observational bounds. Note that the amplitude is still far below possible observations.

\begin{figure}
	\includegraphics[width=\textwidth]{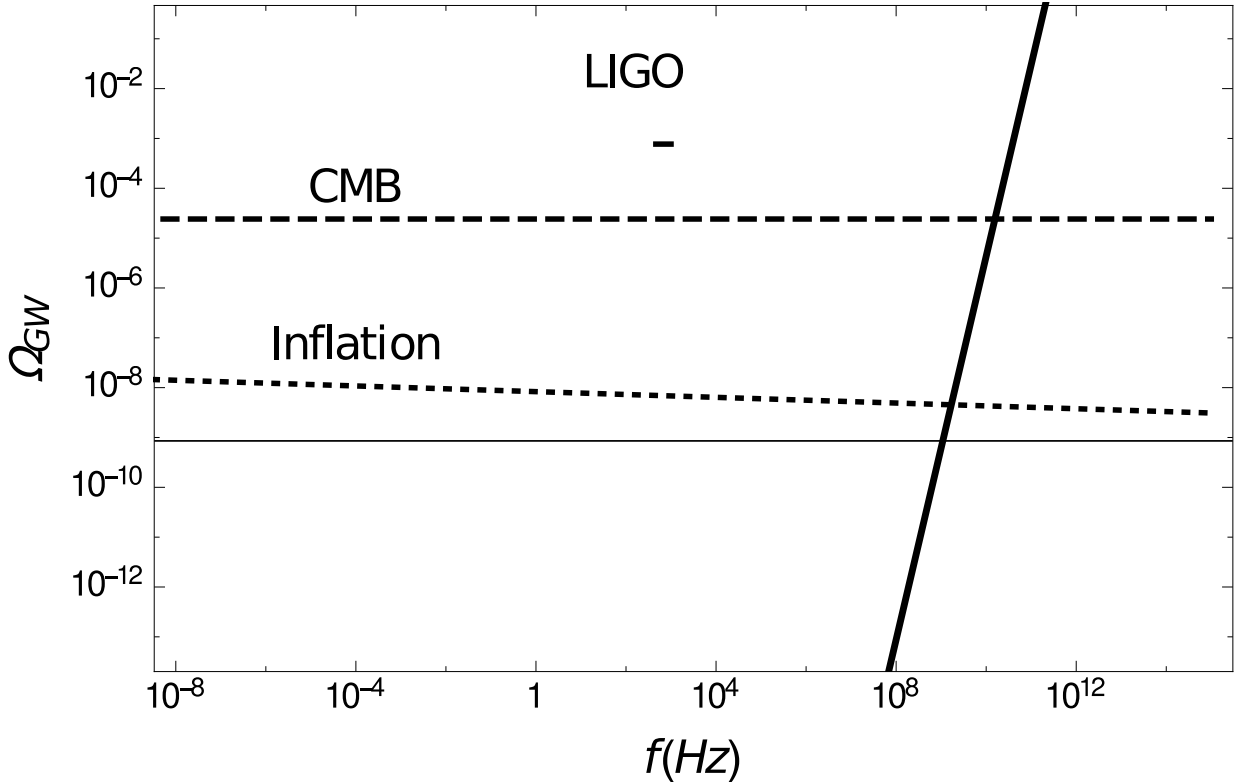}
\caption{\label{fig:High}Amplitude of gravitational waves energy density for high frequencies. The amplitude refers to a model with $a_b=10^{-30}$, and $\lambda=0.9$. The inflation amplitude corresponds to a slow-roll model with $r=T/S=0.6$~\protect\cite{Turner1997}; the CMB line refers to the imprints we should expect in CMB~\protect\cite{Sendra2012}; the LIGO limit is the lower sensitivity for scale-invariant perturbations~\protect\cite{Aasi2015}. Note that the usual flat spacetime ultra-violet divergences were not subtracted in this figure.}
\end{figure}

\end{itemize}

\section{Conclusion}

In bouncing models containing K-essence scalar fields simulating hydrodynamical fluids with $c_s^2=\lambda$,
the amplitude of primordial gravitational waves produced is usually very small~\cite{Bessada2012}
for cosmological scales, or low frequencies, but it can grow significantly at high frequencies
if the fluid which dominates the background dynamics at the bounce is as close
to stiff matter as possible. In this paper, we have shown that this can indeed be true, as one can infer
from Eq.~\eqref{eq:omegaBehave}, but we have also seen that the amplitude of gravitational waves does also
depend on $I_q$ defined on Eq.~\eqref{Iq}, which gets smaller when the bounce fluid approaches stiff matter.
We have seen that the compromise between these two effects makes the amplitude of primordial gravitational waves
not sufficiently big at high frequencies in order to be detected by present day or near future observations
for background models being symmetric around the bounce, and satisfying the nucleosynthesis bounds.
These conclusions are corroborated by
Figs.~(\ref{fig:energydensitychange},\ref{fig:stateparameter},\ref{fig:amplitude4low},\ref{fig:High}),
based on numerical calculations, and understood through analytical considerations. Hence, it seems that
bouncing models where the background is dominated by hydrodynamical fluids do not present any significant
amount of primordial gravitational waves at any frequency range compatible with observations. Any detection
of such waves will then rule out this kind of models.

An alternative would be to consider bouncing models which
are not symmetric around the bounce due, e.g., to particle production near the bounce~\cite{diogo}.
In this case, one could suppose that radiation was created after the bounce, and the nucleosynthesis bounds
originating constraint Eq.~(\ref{eq:omegaconstrain}) could be relaxed, because in the
contracting phase there would be almost
no radiation. It would be a bouncing model with some sort of reheating. In this case, one could have $\lambda$
as close to $1$ as necessary, yielding a sufficiently big $I_q$ as indicated by the $\lambda\approx 1$ part of
Fig.~\ref{fig:omegaint}. In this case, the model could produce a sufficient amount of relic gravitational waves
that could be detected.

\begin{acknowledgments}
N.P.N. would like to thank
CNPq of Brazil for financial support, and A.C.S.
would like to thank FAPERJ of Rio de Janeiro for financial support.
\end{acknowledgments}

\bibliography{bibli}
\end{document}